\documentclass[aps,preprint,showpacs]{revtex4}
\usepackage{epsf}
\usepackage{epsfig}
\usepackage{amsmath}

\def\beq{\begin{equation}}
\def\eeq{\end{equation}}
\def\bea{\begin{eqnarray}}
\def\eea{\end{eqnarray}}
\def\beqa{\begin{equation}\begin{array}{l}}
\def\eeqa{\end{array}\end{equation}}

\begin{document}

\title{
Empirical transverse charge densities in the nucleon-to-$P_{11}(1440)$
transition}

\author{Lothar Tiator}
\author{Marc Vanderhaeghen}

\affiliation{Institut f\"ur Kernphysik, Johannes Gutenberg-Universit\"at,
D-55099 Mainz, Germany}

\date{\today}

\begin{abstract}
Using recent experimental data, we analyze the electromagnetic
transition from the nucleon to the $P_{11} (1440)$ resonance. From
the resulting empirical transition form factors, we map out the
quark transverse charge densities which induce the $N \to P_{11}
(1440)$ transition. It is found that the transition from the proton
to its first radially excited state is dominated by up quarks in a
central region of around 0.5~fm and by down quarks in an outer band
which extends up to about 1~fm.
\end{abstract}

\pacs{13.40.Gp, 13.40.Em, 14.20.Gk}

\maketitle
\thispagestyle{empty}

A major focuss of current research in hadronic physics centers
around the question of how the structure of the nucleon and its
excitations can be quantitatively understood from the interaction
among its constituent quarks and gluons. To this end, a substantial
experimental effort is underway at electron facilities such as JLab,
ELSA, and MAMI to map out the nucleon excitation spectrum and to
measure the electromagnetic transition form factors from a nucleon
to an excited baryon.

The form factors (FFs) describing the electromagnetic (e.m.)
transition of the nucleon to the first baryon excitation, the
$\Delta(1232)$ resonance, have been measured precisely and over a
large range of photon virtualities by means of the $e p \to e p
\pi^0$ and $e p \to e n \pi^+$ reaction. The resulting data allow to
study and quantify the deformation of the $N \to \Delta$ transition
charge distribution, see e.g. Ref.~\cite{Pascalutsa:2006up} for a
recent review and references therein.

High precision data have also become available in recent years for
the $\gamma^\ast N \to N^\ast$ transition, for the
$P_{11}(1440)$~\cite{MAID2003,Aznauryan:2004jd,MAID2007,Aznauryan:2008pe},
$D_{13}(1520)$~\cite{MAID2003,Aznauryan:2004jd,MAID2007},
$S_{11}(1535)$~\cite{MAID2003,Aznauryan:2004jd,Denizli:2007tq,MAID2007}
and $F_{15}(1680)$~\cite{MAID2003,MAID2007} nucleon resonances, from
$\pi^0$~\cite{Joo:2005gs,Biselli:2008zz},
$\pi^+$~\cite{Joo:2004mi,Egiyan:2006ks,Park:2007tn}, and
$\eta$~\cite{Armstrong:1998wg,Thompson:2000by,Denizli:2007tq,Merkel:2007ig}
electroproduction data on the nucleon.

On the theoretical side, the nucleon-to-resonance transition FFs are
also becoming amenable to lattice QCD calculations. For the $N \to
\Delta$ transition, first full QCD results using different fermion
actions were presented in Ref.~\cite{Alexandrou:2007dt}. It was
found that the unquenched results for the small $N \to \Delta$
Coulomb quadrupole FF show deviations from the unquenched one at low
$Q^2$, underlining the important role of the pion cloud to this
observable. For the electromagnetic transition of the nucleon to its
first excited state with quantum numbers $J^P = \frac{1}{2}^+$, the
$P_{11}(1440)$ resonance, often referred to as the {\it Roper}
resonance, first full QCD lattice studies were performed in
Ref.~\cite{Lin:2008qv}. The Roper resonance, being the first radial
excitation of the nucleon, presents a particular challenge for
lattice calculations to reproduce the correct level ordering with
its negative-parity partner, the $S_{11}(1535)$ resonance.

The precise e.m. FF data, extracted from experiment, allow to map
out the quark charge densities in a baryon. It was shown possible to
define a proper density interpretation of the form factor data by
viewing the baryon in a light-front frame. This yields information
on the spatial distribution of the quark charge in the plane
transverse to the line-of-sight. In this way, the quark transverse
charge densities were mapped out in the
nucleon~\cite{Miller:2007uy,Carlson:2007xd}, and in the
deuteron~\cite{Carlson:2008zc} based on empirical FF data.
Furthermore, recent lattice QCD results were used to map out the
quark transverse densities in the $\Delta
(1232)$~\cite{Alexandrou:2008bn} resonance. To understand the e.m.
structure of a nucleon resonance, it is of interest to use the
precise transition FF data to reveal the spatial distribution of the
quark charges that induce such a transition. In this way, the $N \to
\Delta(1232)$ transition charge densities have been mapped out in
Ref.~\cite{Carlson:2007xd} using the empirical information of the $N
\to \Delta(1232)$ transition FFs~\cite{MAID2007}. In the following,
we will generalize the above considerations to the e.m. transition
between the nucleon and the first excited state with nucleon quantum
numbers, the $P_{11}(1440)$ resonance. We will use the empirical
information to map out the quark transition charge densities
inducing the $N \to P_{11}(1440)$ e.m. excitation.

After outlining the definitions to characterize the vertex for the
$N \to P_{11}(1440)$ e.m. transition, we will use the recent
JLab/CLAS data to extract the two independent $N \to P_{11}(1440)$
transition FFs. Subsequently, we define the transition densities and
use the empirical FF information to map out the spatial distribution
of quark charges that induce the $N \to P_{11} (1440)$ transition.
We finally give a brief summary and outline extensions of this work.

Consider the e.m. transition from the nucleon to its first excited
state, the $P_{11}(1440)$ resonance, which we will denote in the
following by $N^\ast$. A Lorentz-covariant decomposition of the
matrix element of the electromagnetic current operator $J^\mu$ for
this transition, satisfying manifest electromagnetic
gauge-invariance, can be written as~:
\begin{eqnarray}
\langle N^\ast(p',\lambda^\prime) \,|\,J^\mu(0) \,|\, N (p,\lambda)
\rangle
&=& \bar u (p',\lambda^\prime) \left\{
F_1^{N N^\ast}(Q^2)
\left( \gamma^\mu - \gamma \cdot q \, \frac{q^\mu}{q^2} \right)
\right. \nonumber \\
&&\left. \hspace{1.5cm}
+ F_2^{N N^\ast}(Q^2) \frac{i \sigma^{\mu\nu} q_\nu}{(M^\ast + M_N)} \,
\right\} u (p,\lambda) ,
\label{eq:nnstartree}
\end{eqnarray}
where $M_N$ is the nucleon mass, $M^\ast = 1.440$~GeV is the mass of
the first excited resonance with nucleon quantum numbers, $\lambda$
($\lambda^\prime$) are the initial (final) baryon helicities, and
$u$ is the spin 1/2 spinor, normalized as $\bar u u = 2 M$.
Furthermore, $F^{N N^\ast}_{1,2}$ are the electromagnetic (e.m.) FFs
for the $N \to N^\ast$ transition.

Equivalently, one can also parametrize the $\gamma^\ast N N^\ast$
transition through two helicity amplitudes $A_{1/2}$ and
$S_{1/2}$, which are defined in the $N^\ast$ rest frame.
These $N^\ast$ rest frame helicity amplitudes are defined
through the following matrix elements of the electromagnetic current
operator:
\begin{eqnarray}
A_{1/2} \,&\equiv&\, -\frac{e}{\sqrt{2 K}} \;
\frac{1}{(4 M_N M^\ast)^{1/2}} \; \langle \; N^\ast(\vec 0, \, +1/2)
\,|\, {\bf J \cdot \epsilon}_{\lambda = +1} \,|\, N(-\vec q, \, -1/2 )
\;\rangle,  \nonumber \\
S_{1/2} \,&\equiv&\, \frac{e}{\sqrt{2 K}} \;
\frac{1}{(4 M_N M^\ast)^{1/2}} \; \langle \; N^\ast(\vec 0, \, +1/2)
\,|\, J^0 \,|\, N(-\vec q, \, +1/2 )
\;\rangle,
\label{eq:resthel}
\end{eqnarray}
where the spin projections are along the $z$-axis (chosen along the
virtual photon direction) and where the transverse photon
polarization vector entering $A_{1/2}$ is given by ${\bf
\epsilon}_{\lambda = +1} = -1/\sqrt{2} (1, i, 0)$. Furthermore in
Eq.~(\ref{eq:resthel}), $e$ is the proton electric charge, related
to the fine-structure constant as $\alpha_{em} \equiv e^2 /(4 \pi)
\simeq 1/137$, and $K$ is the ``equivalent photon energy'' defined
as~:
\begin{eqnarray}
K \,\equiv \, \frac{M^{\ast \, 2} - M_N^2}{2 M^\ast} .
\label{eq:equivga}
\end{eqnarray}
The helicity amplitudes are functions of the photon virtuality $Q^2$,
and can be expressed in terms of the FFs $F_1^{N N^\ast}$
and $F_2^{N N^\ast}$ as~:
\begin{eqnarray}
A_{1/2} &=& e \, \frac{Q_-}{\sqrt{K} \, (4 M_N M^\ast)^{1/2}}
\left\{ F_1^{N N^\ast}  + F_2^{N N^\ast} \right\} , \\
S_{1/2} &=& e \, \frac{Q_-}{\sqrt{2 K} \, (4 M_N M^\ast)^{1/2}}
\, \left( \frac{Q_+ Q_-}{2 M^\ast} \right) \,
\frac{(M^\ast + M_N)}{Q^2}  \left\{ F_1^{N N^\ast} -
\frac{Q^2}{(M^\ast + M_N)^2} F_2^{N N^\ast} \right\} ,
\label{eq:helff}
\end{eqnarray}
where we introduced the shorthand notation
$Q_\pm \equiv \sqrt{(M^\ast \pm M_N)^2 +Q^2}$.

For numerical evaluation, we will use the MAID2007
parameterization~\cite{MAID2007}
for the $\gamma^* N \to P_{11}(1440)$
helicity amplitudes $A_{1/2}$ and $S_{1/2}$.
They have been parameterized as~:
\begin{eqnarray}
A_{1/2}(Q^2) &=& A^0_{1/2} \, \left( 1 + a_1 \, Q^2 + a_2 \, Q^4 +
a_3 \, Q^8 \right)
e^{- a_4 Q^2} , \\
S_{1/2}(Q^2) &=& S^0_{1/2} \, \left( 1 + s_1 \, Q^2 + s_2 \, Q^4 +
s_3 \, Q^8 \right) e^{- s_4 Q^2} .
\end{eqnarray}
The improved proton fit is based on $\pi^0 p$ data
from~\cite{Frolov:1998pw,Joo:2005gs,Ungaro:2006df} and $\pi^+ n$
data from~\cite{Joo:2004mi,Egiyan:2006ks,Park:2007tn}. The neutron
fit is based only an older pre-2000 quasi-free $\pi^- p$ data from
the SAID data base~\cite{SAID}. The resulting values of the
parameters are given in Table~(\ref{table:maidfitA},
\ref{table:maidfitS}) for both proton and neutron.

\begin{table}[ht]
\begin{center}
\begin{tabular}{|c|c|c|c|c|c|}
\hline
\hline
& $A^0_{1/2}$ & $a_1$   & $a_2$ & $a_3$& $a_4$      \\
& ($10^{-3}$ GeV$^{-1/2}$) & (GeV$^{-2}$) & (GeV$^{-4}$)
&(GeV$^{-8}$) & (GeV$^{-2}$)  \\
\hline \;\; p  \;\; &
$-61.4$ & $ 0.871 $ & $-3.52$ &-0.158& 1.36  \\
\hline \;\; n \;\; & 54.1 & 0.95 & 0 & 0 & 1.77 \\
\hline
\hline
\end{tabular}
\end{center}
\caption{Parameters for the transverse $\gamma^* N \to P_{11}(1440)$
helicity amplitudes for the proton (first row) and for the neutron
(second row), according to the MAID2007 fit~\cite{MAID2007} for the neutron,
and the improved MAID2007 fit for the proton.}
\label{table:maidfitA}
\end{table}%

\begin{table}[ht]
\begin{center}
\begin{tabular}{|c|c|c|c|c|c|}
\hline
\hline
& $S^0_{1/2}$ & $s_1$   & $s_2$ & $s_3$& $s_4$      \\
& ($10^{-3}$ GeV$^{-1/2}$) & (GeV$^{-2}$) & (GeV$^{-4}$)
&(GeV$^{-8}$) & (GeV$^{-2}$)  \\
\hline \;\; p \;\; &
4.2 & 40.0 & 0 & 1.5 & 1.75   \\
\hline \;\; n \;\; & -41.5 & 2.98 & 0 & 0 & 1.55 \\
\hline
\hline
\end{tabular}
\end{center}
\caption{Parameters for the longitudinal $\gamma^* N \to
P_{11}(1440)$ helicity amplitudes for the proton (first row) and for
the neutron (second row), according to the MAID2007 fit for the neutron,
and the improved MAID2007 fit for the proton.}
\label{table:maidfitS}
\end{table}%

\begin{figure}
\begin{center}
\includegraphics[width =5.cm, angle=90]{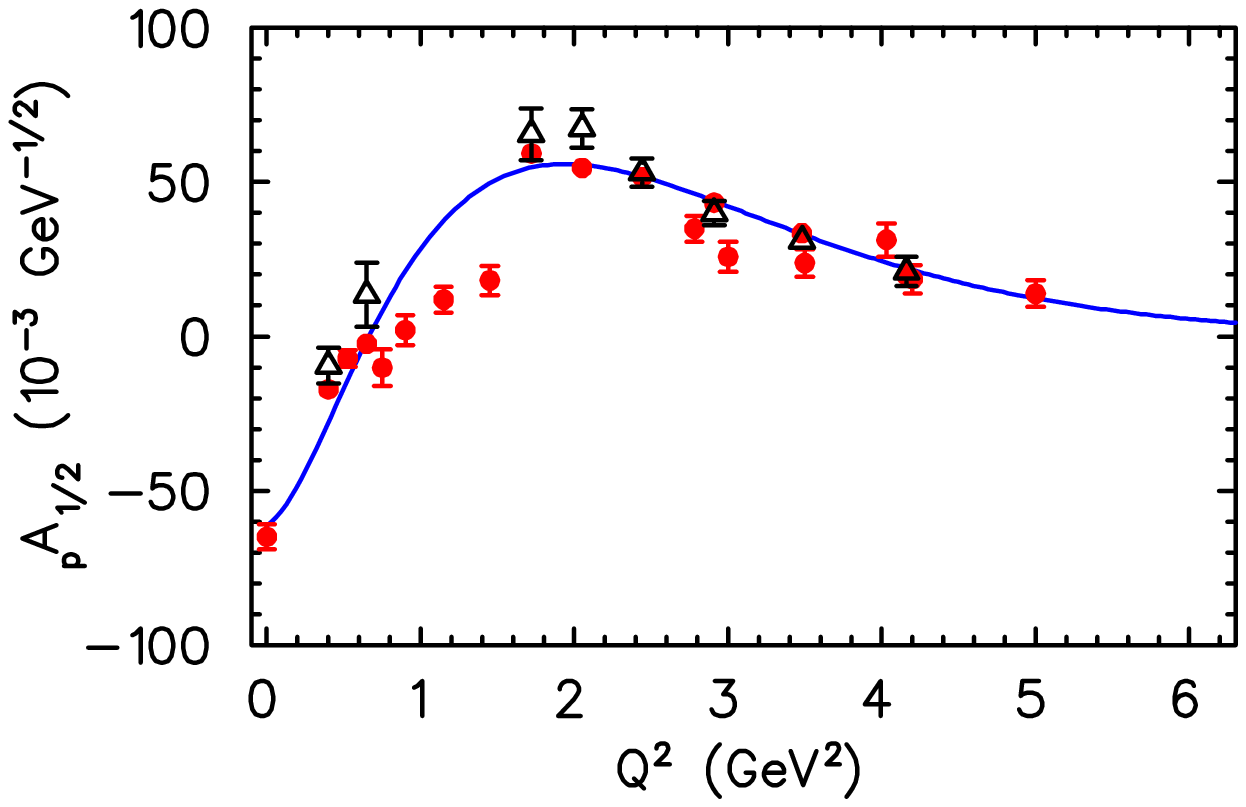}
\hspace{0.5cm}
\includegraphics[width =5.cm, angle=90]{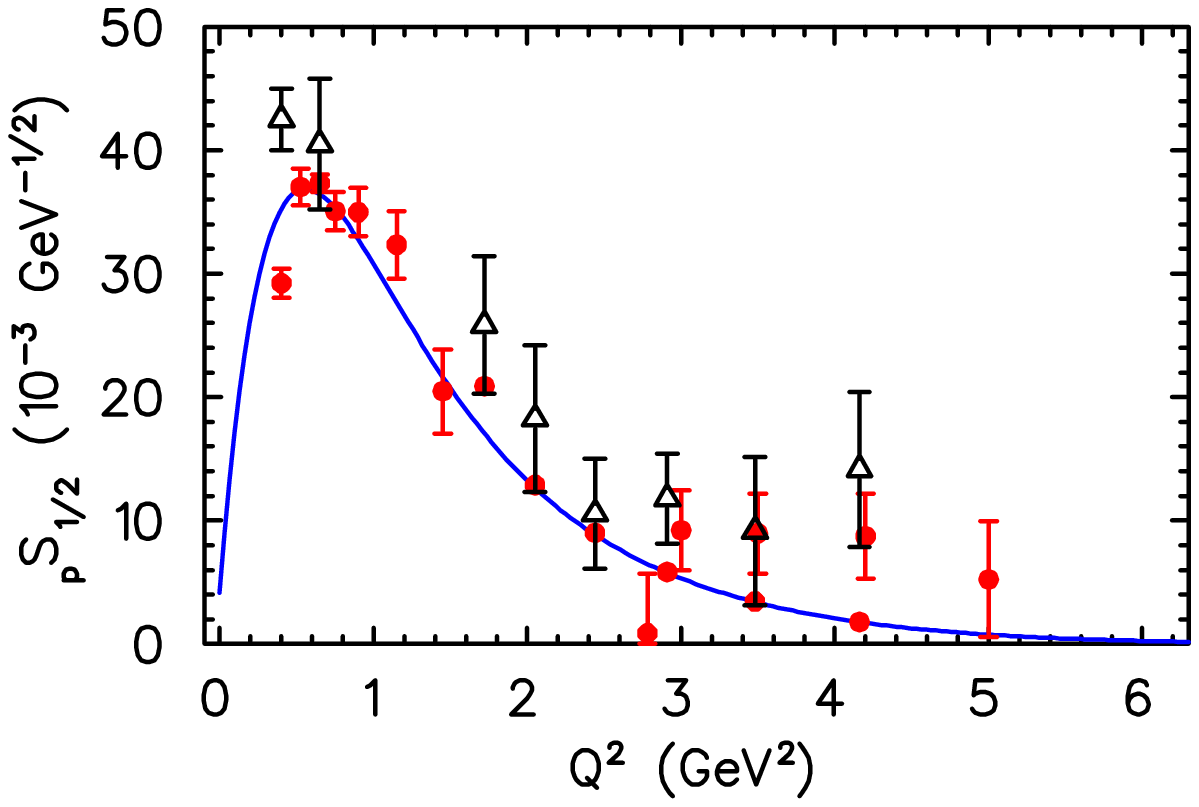}
\end{center}
\vspace{-1cm}
\caption{Helicity amplitudes for the $\gamma^* p \to P_{11}(1440)$ transition.
The solid circles are the improved MAID2007 analysis from this work.
The open triangles are the analysis of
Refs.~\cite{Aznauryan:2004jd,Aznauryan:2008pe}.
The point at $Q^2 = 0$ is from the PDG~\cite{Amsler:2008zz}.
}
\label{fig:nnstar1}
\end{figure}

In Fig.~\ref{fig:nnstar1}, we show the helicity amplitudes for the
$\gamma^* p \to P_{11}(1440)$ transition, which have been measured
up to $Q^2 \simeq 5$~GeV$^2$. One notices that the helicity
amplitude $A_{1/2}$ for transverse photons displays a sign change
from a large negative value at the real photon point to a broad
positive maximum around $Q^2 \simeq$~2~GeV$^2$. The helicity
amplitude $S_{1/2}$ for longitudinal photons stays positive and has
a maximum around $Q^2 \simeq$~0.6~GeV$^2$.

\begin{figure}
\begin{center}
\includegraphics[width =5.cm, angle=90]{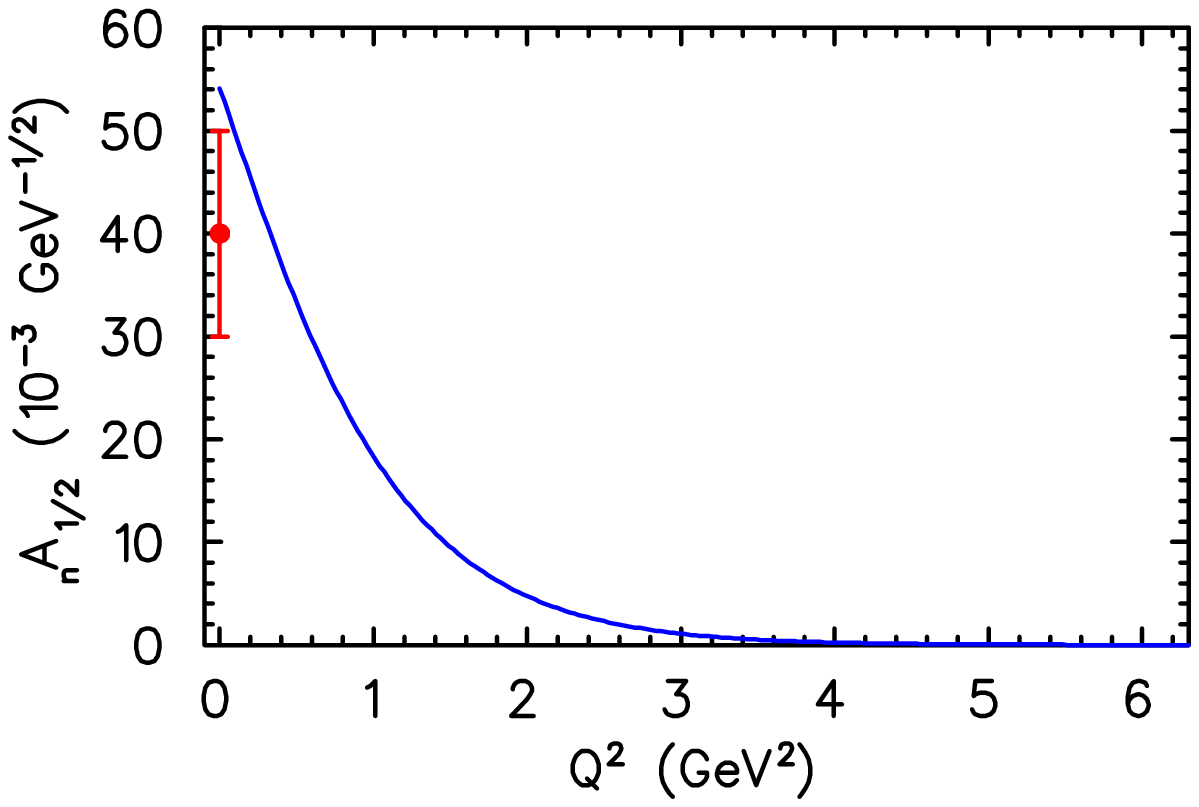}
\hspace{0.5cm}
\includegraphics[width =5.cm, angle=90]{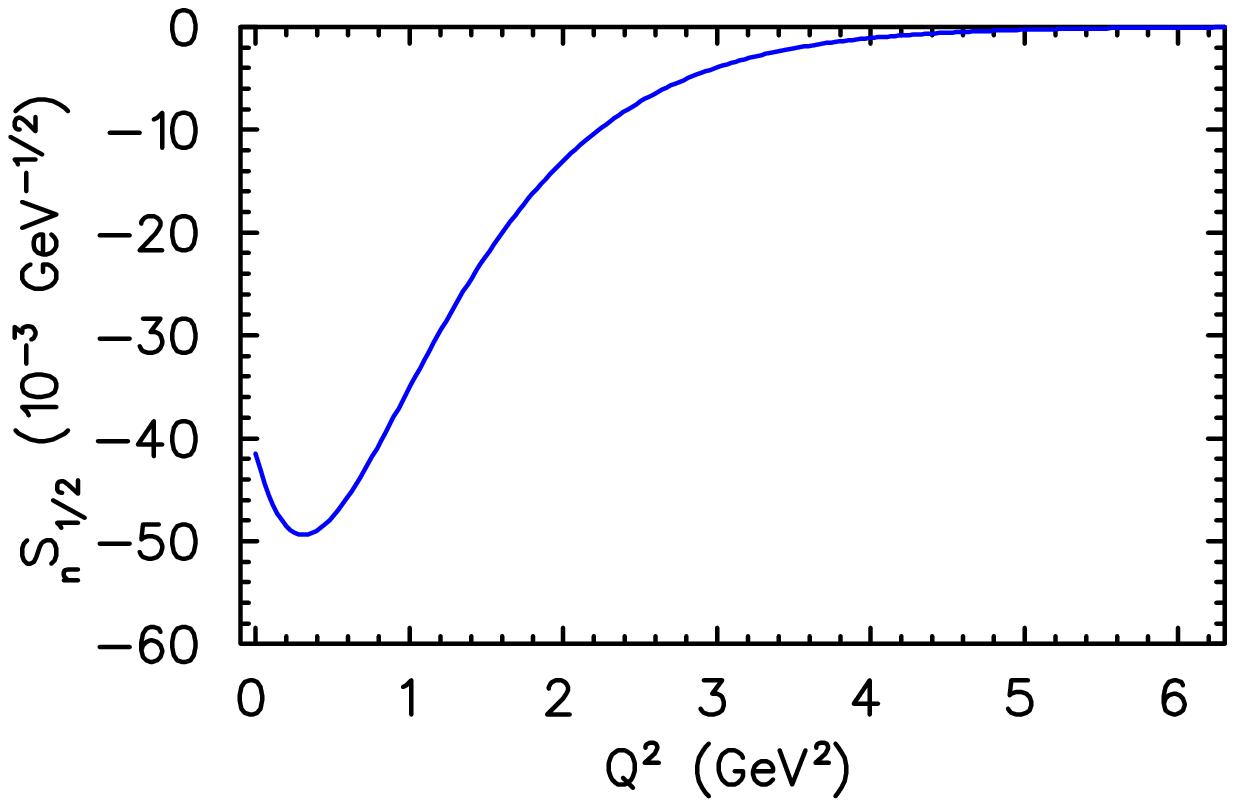}
\end{center}
\vspace{-1cm}
\caption{Helicity amplitudes for the $\gamma^* n \to P_{11}(1440)$ transition,
 according to the MAID2007~\cite{MAID2007} analysis.
The point at $Q^2 = 0$ is from the PDG~\cite{Amsler:2008zz}.
}
\label{fig:nnstar2}
\end{figure}

The corresponding helicity amplitudes for the neutron are shown in
Fig.~\ref{fig:nnstar2}. One sees that apart from the value at the real
photon point for $A_{1/2}$, these amplitudes are yet to be measured.
The MAID2007 analysis shows an $A_{1/2}$ helicity amplitude for the neutron
which does not display a sign change as in the proton case.

\begin{figure}
\begin{center}
\includegraphics[width =5.cm, angle=90]{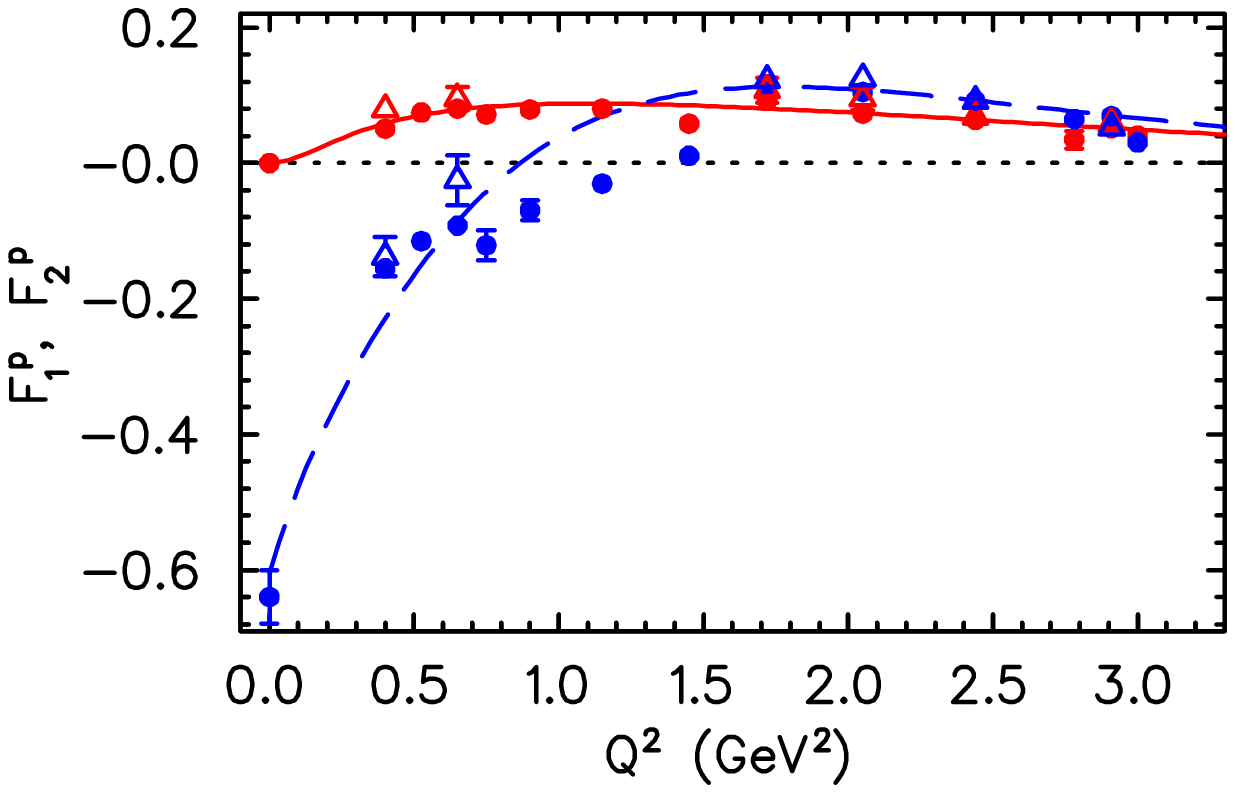}
\hspace{0.0cm}
\includegraphics[width =5.cm, angle=90]{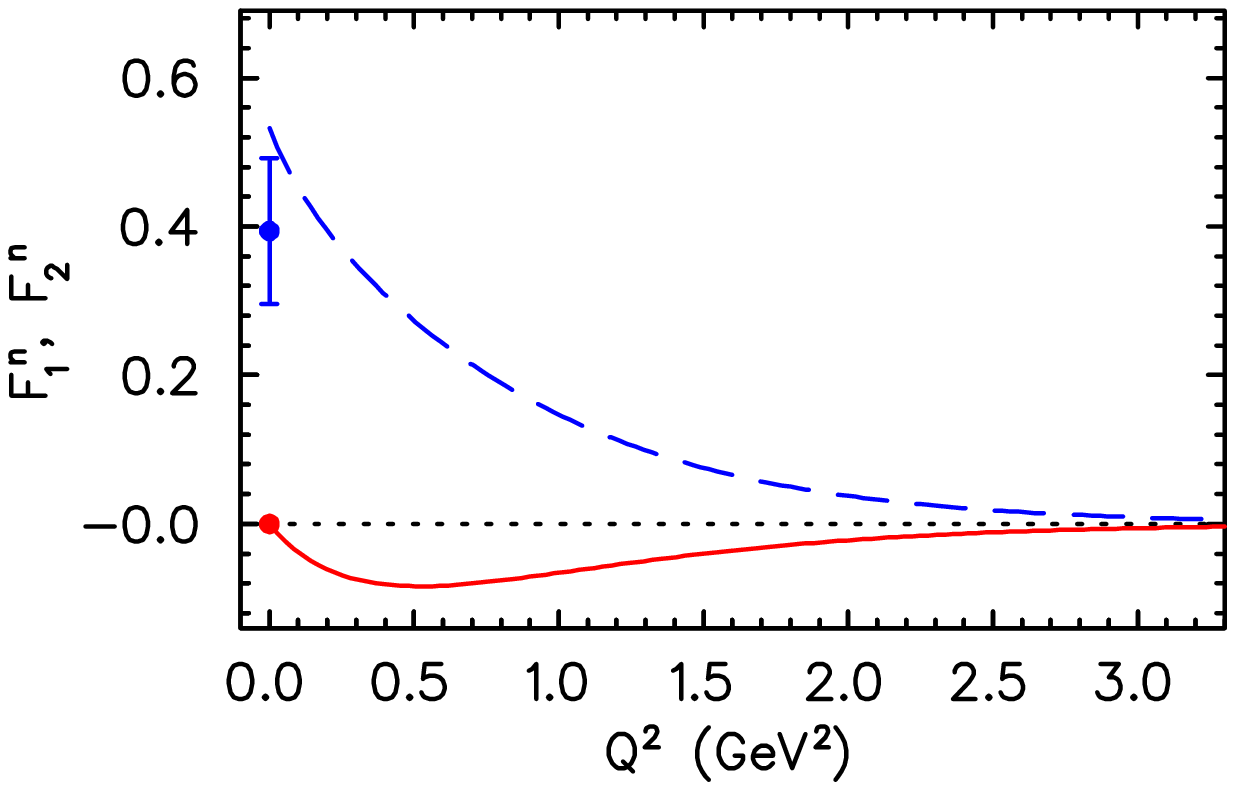}
\end{center}
\vspace{-1cm}
\caption{Form factors for the $\gamma^* N \to P_{11}(1440)$ transition,
for the proton (left panel) and the neutron (right panel),
based on the MAID2007~\cite{MAID2007} analysis for the neutron and
the improved MAID2007 analysis from this work.
Solid curves : $F^{N N^\ast}_1$; dashed curves : $F^{N N^\ast}_2$.}
\label{fig:nnstar3}
\end{figure}

In Fig.~\ref{fig:nnstar3}, we show the MAID analysis for the
$\gamma^* N \to P_{11} (1440)$ transition form factors defined in
Eq.~(\ref{eq:nnstartree}). They are related with the helicity
amplitudes as in Eq.~(\ref{eq:helff}). One sees that for the proton,
the Dirac-like FF $F_1^{N N^\ast}$ vanishes with $Q^2$ when
approaching the real photon point and stays positive at large $Q^2$.
On the other hand, the Pauli-type FF  $F_2^{N N^\ast}$ assumes a
large negative value at the real photon point, $F_2^{p N^\ast}(0) =
-0.64 \pm 0.04$, and changes sign around $Q^2 \simeq 1$~GeV$^2$.

In the following, we will consider the e.m. $N \to P_{11}(1440)$
transition when viewed from a light front moving towards the baryon.
Equivalently, this corresponds with a frame where the baryons have a
large momentum-component along the $z$-axis chosen along the
direction of $P = (p + p^\prime)/2$, where $p$ ($p^\prime$) are the
initial (final) baryon four-momenta. We indicate the baryon
light-front + component by $P^+$ (defining $a^\pm \equiv a^0 \pm
a^3$). We can furthermore choose a symmetric frame where the virtual
photon four-momentum $q$ has $q^+ = 0$, and has a transverse
component (lying in the $xy$-plane) indicated by the transverse
vector $\vec q_\perp$, satisfying $q^2 = - {\vec q_\perp}^{\, 2}
\equiv - Q^2$. In such a symmetric frame, the virtual photon only
couples to forward moving partons and the + component of the
electromagnetic current $J^+$ has the interpretation of the quark
charge density operator. It is given by~: $J^+(0) = +2/3 \, \bar
u(0) \gamma^+ u(0) - 1/3 \, \bar d(0) \gamma^+ d(0)$, considering
only $u$ and $d$ quarks. Each term in the expression is a positive
operator since $\bar q \gamma^+ q \propto | \gamma^+ q |^2$.

We define a transition charge density for the unpolarized
$N \to N^\ast$ transition, by the Fourier transform~:
\begin{eqnarray}
\rho_0^{N N^\ast}(\vec b) &\equiv& \int \frac{d^2 \vec q_\perp}{(2 \pi)^2} \,
e^{- i \, \vec q_\perp \cdot \vec b} \, \frac{1}{2 P^+}
\langle P^+, \frac{\vec q_\perp}{2}, \lambda \,|\, J^+(0) \,|\,
P^+, -\frac{\vec q_\perp}{2}, \lambda  \rangle ,
\label{eq:ndens0}
\end{eqnarray}
where $\lambda$ ($\lambda^\prime$) denotes the nucleon ($N^\ast$)
light-front helicities,
$\vec q_\perp = Q ( \cos \phi_q \hat e_x + \sin \phi_q \hat e_y )$,
and where the 2-dimensional vector $\vec b$ denotes the position (in the
$xy$-plane) from the transverse {\it c.m.} of the baryons.
The Fourier transform in Eq.~(\ref{eq:ndens0}) can be worked out as~:
\begin{eqnarray}
\rho_0^{N N^\ast}(\vec b) =
\int_0^\infty \frac{d Q}{2 \pi} Q \, J_0(b \, Q) F_1^{N N^\ast}(Q^2),
\label{eq:ndens1}
\end{eqnarray}
where $J_n$ denotes the cylindrical Bessel function of order $n$.
Note that $\rho_0^{N N^\ast}$ only depends on $b = |\vec b|$.
It has the interpretation of the quark (transition) charge density in the
transverse plane which induces the $N \to N^\ast$ excitation.

The above unpolarized transition charge density involves only one of
the two independent $N \to N^\ast$ e.m. FFs. To extract the information
encoded in $F_2^{N N^\ast}$, we consider the transition charge densities
for a transversely polarized $N$ and $N^\ast$.
We denote this transverse polarization direction by
$\vec S_\perp = \cos \phi_S \hat e_x + \sin \phi_S \hat e_y$.
The transverse spin state can be expressed in terms of the light
front helicity spinor states as~:
$| s_\perp = + \frac{1}{2} \rangle = \left( | \lambda = + \frac{1}{2} \rangle
+ e^{i \phi_S } \, | \lambda = - \frac{1}{2} \rangle \right) / \sqrt{2}$,
with $s_\perp$ the nucleon
spin projection along the direction of $\vec S_\perp$.

We can then define a transition charge density for a
transversely polarized $N$ and $N^\ast$,
both along the direction of $\vec S_\perp$ as~:
\begin{eqnarray}
\rho_T^{N N^\ast}(\vec b) &\equiv& \int \frac{d^2 \vec q_\perp}{(2 \pi)^2} \,
e^{-i \, \vec q_\perp \cdot \vec b} \, \frac{1}{2 P^+}
\langle P^+, \frac{\vec q_\perp}{2}, s_\perp
\,|\, J^+(0) \,|\,
P^+, -\frac{\vec q_\perp}{2}, s_\perp   \rangle.
\label{eq:ndens2}
\end{eqnarray}
Using Eqs.~(\ref{eq:nnstartree}) and~(\ref{eq:ndens1}),
the Fourier transform of Eq.~(\ref{eq:ndens2}) can be worked out as~:
\begin{eqnarray}
\rho_T^{N N^\ast}(\vec b) &=& \rho_0^{ N N^\ast}(b)
+ \sin (\phi_b - \phi_S) \,
\int_0^\infty \frac{d Q}{2 \pi} \frac{Q^2}{(M^\ast + M_N)} \, J_1(b \, Q)
F_2^{N N^\ast}(Q^2),
\label{eq:ndens3}
\end{eqnarray}
where the second term, which describes the deviation from the circular
symmetric unpolarized charge density, depends on the orientation of
$\vec b = b ( \cos \phi_b \hat e_x + \sin \phi_b \hat e_y )$.
In the following we choose the transverse spin along the $x$-axis
($\Phi_S = 0$).

\begin{figure}
\begin{center}
\includegraphics[width =8.cm]{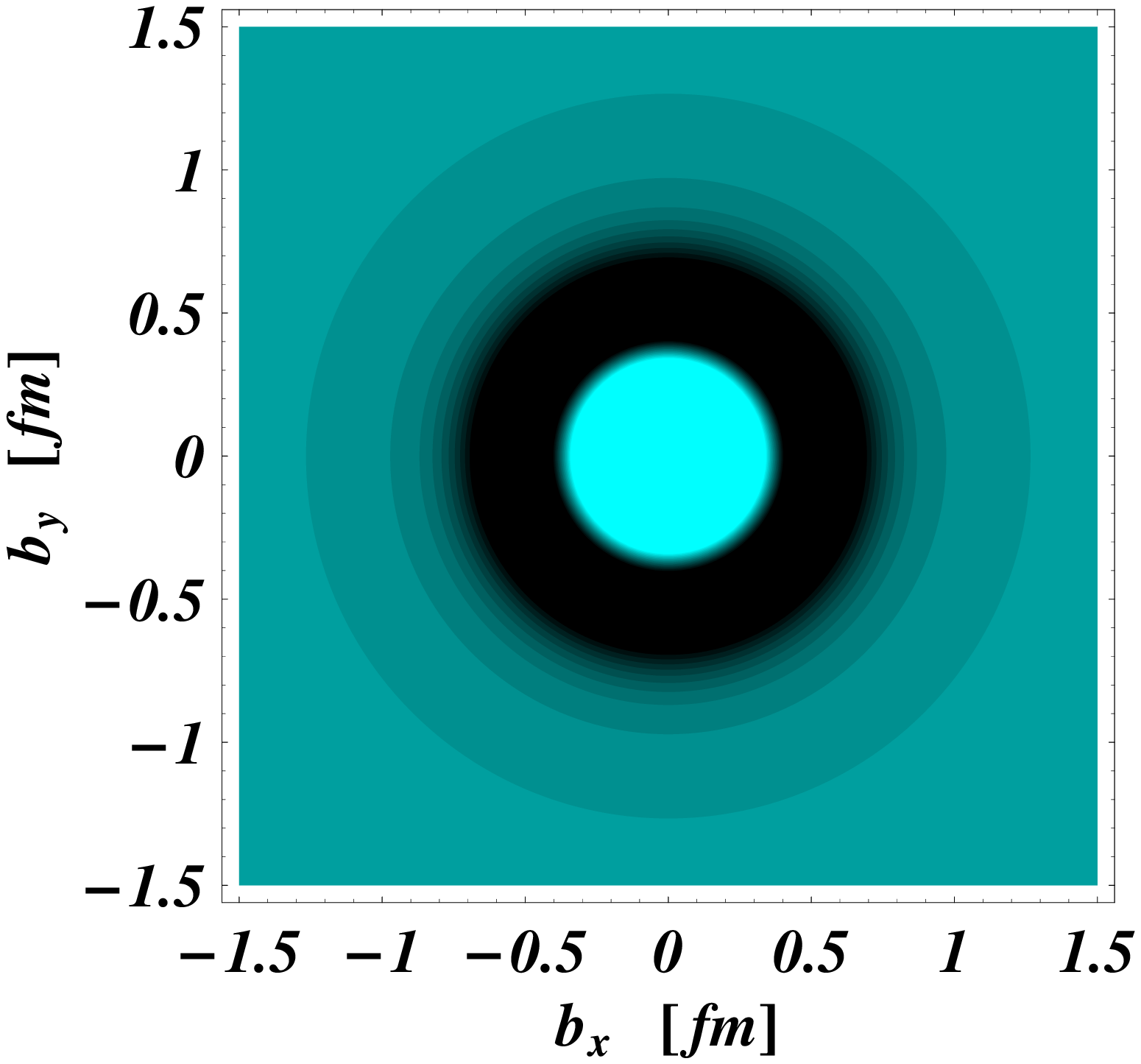}
\includegraphics[width =8.cm]{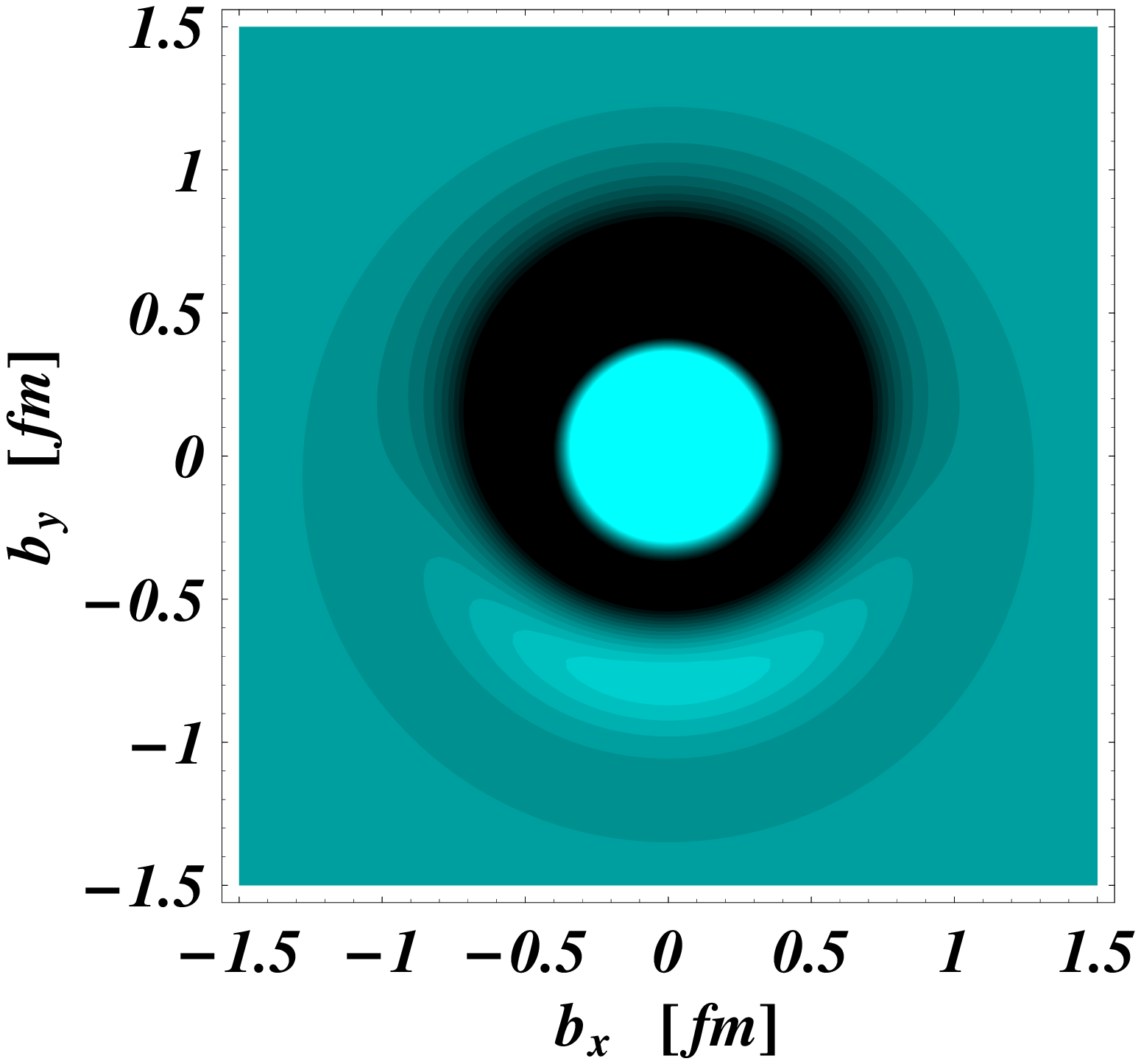}
\vspace{1.0cm}

\includegraphics[width =6.cm, angle=90]{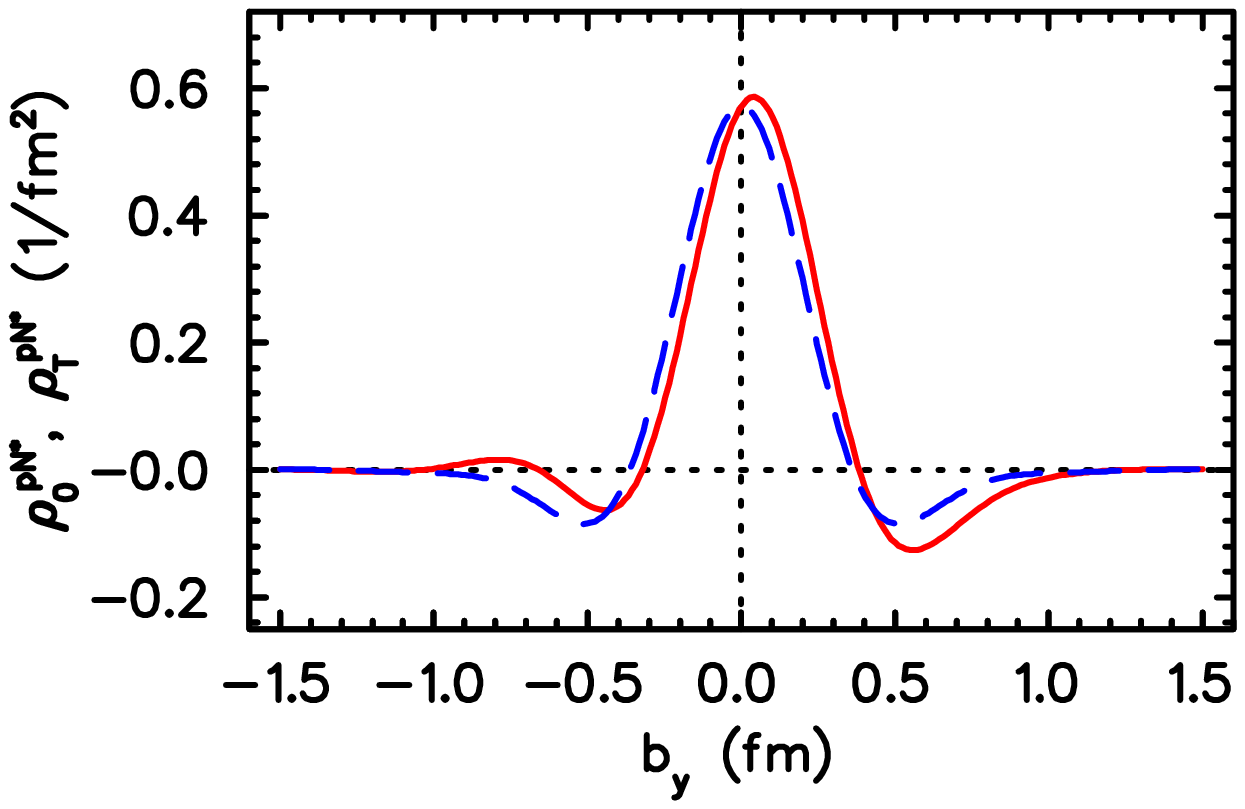}
\end{center}
\caption{Quark transverse charge density
corresponding to the $p \to P_{11}(1440)$ e.m. transition.
Upper left panel : when $p$ and $N^\ast$
are unpolarized ($\rho_0^{p N^\ast}$).
Upper right panel : when $p$ and $N^\ast$ are polarized along the $x$-axis
($\rho_T^{p N^\ast}$). 
The light (dark) regions correspond with positive (negative) densities. 
Lower panel : densities
$\rho_T^{p N^\ast}$ (solid curves) and
$\rho_0^{p N^\ast}$ (dashed curves) along the $y$-axis.
For the $p \to P_{11}(1440)$ e.m. transition FFs,
we use the improved MAID2007 fit of this work. }
\label{fig:nnstar4}
\end{figure}
\begin{figure}
\begin{center}
\includegraphics[width =8.cm]{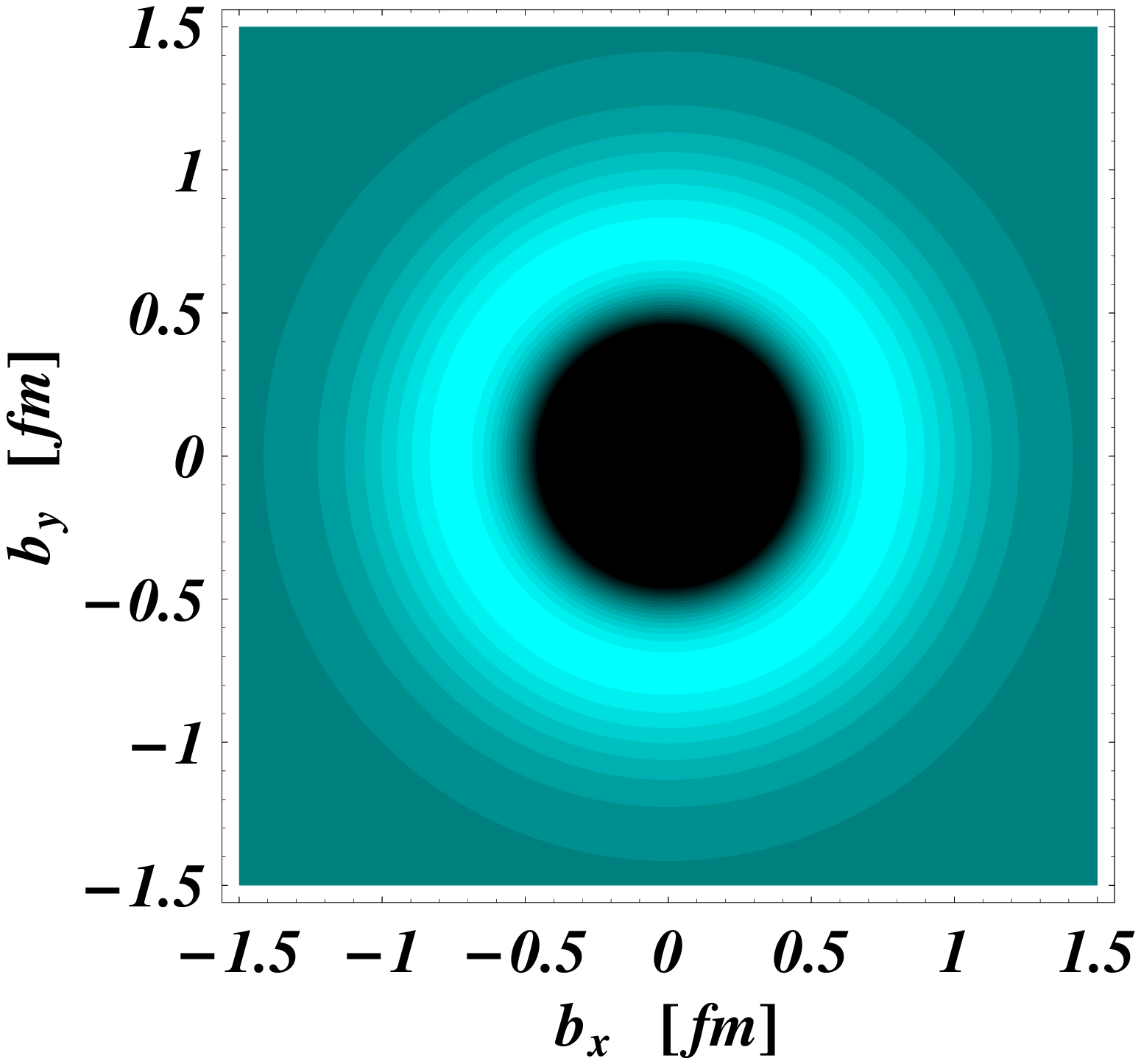}
\includegraphics[width =8.cm]{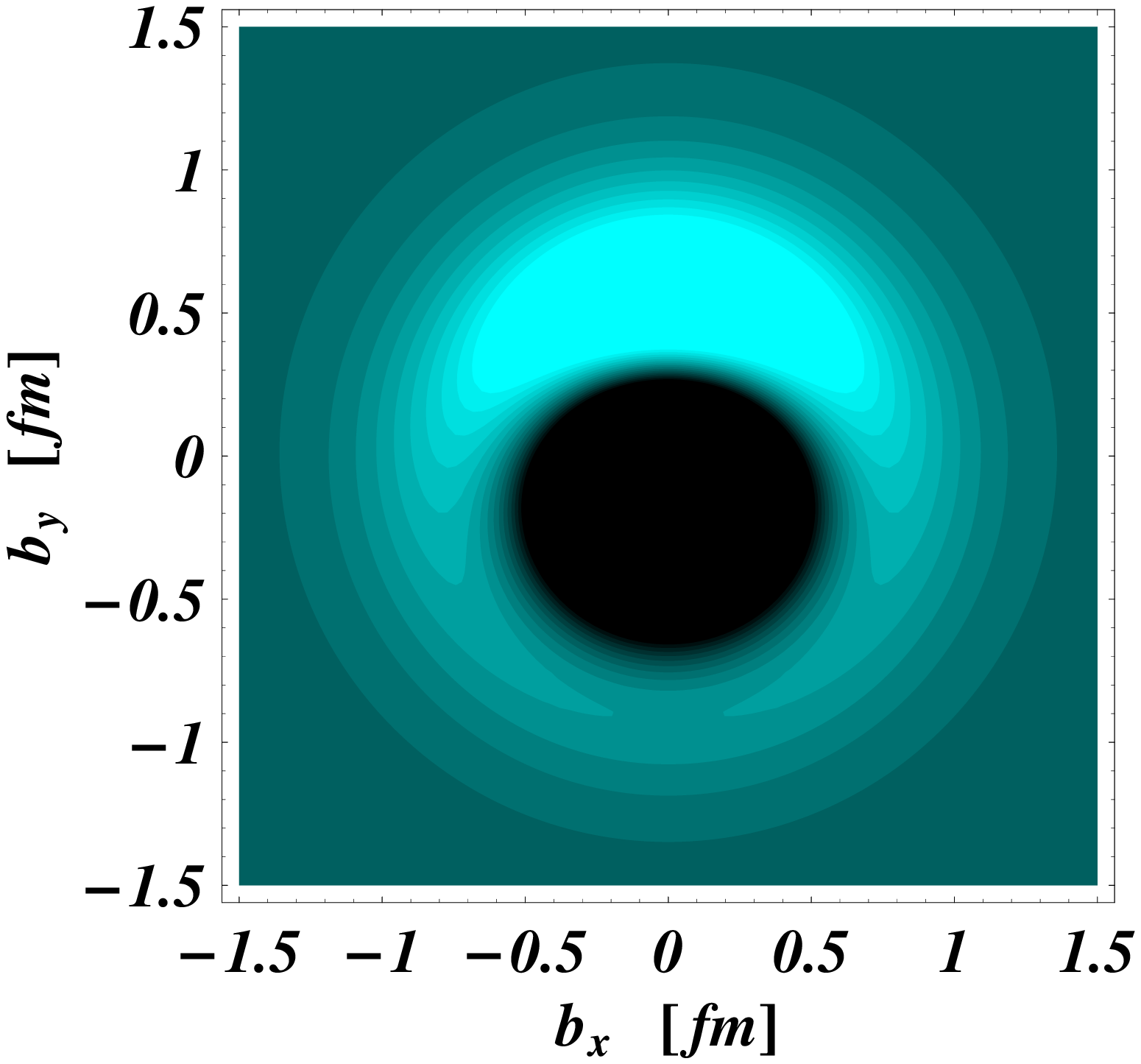}
\vspace{1.0cm}

\includegraphics[width =6.cm, angle=90]{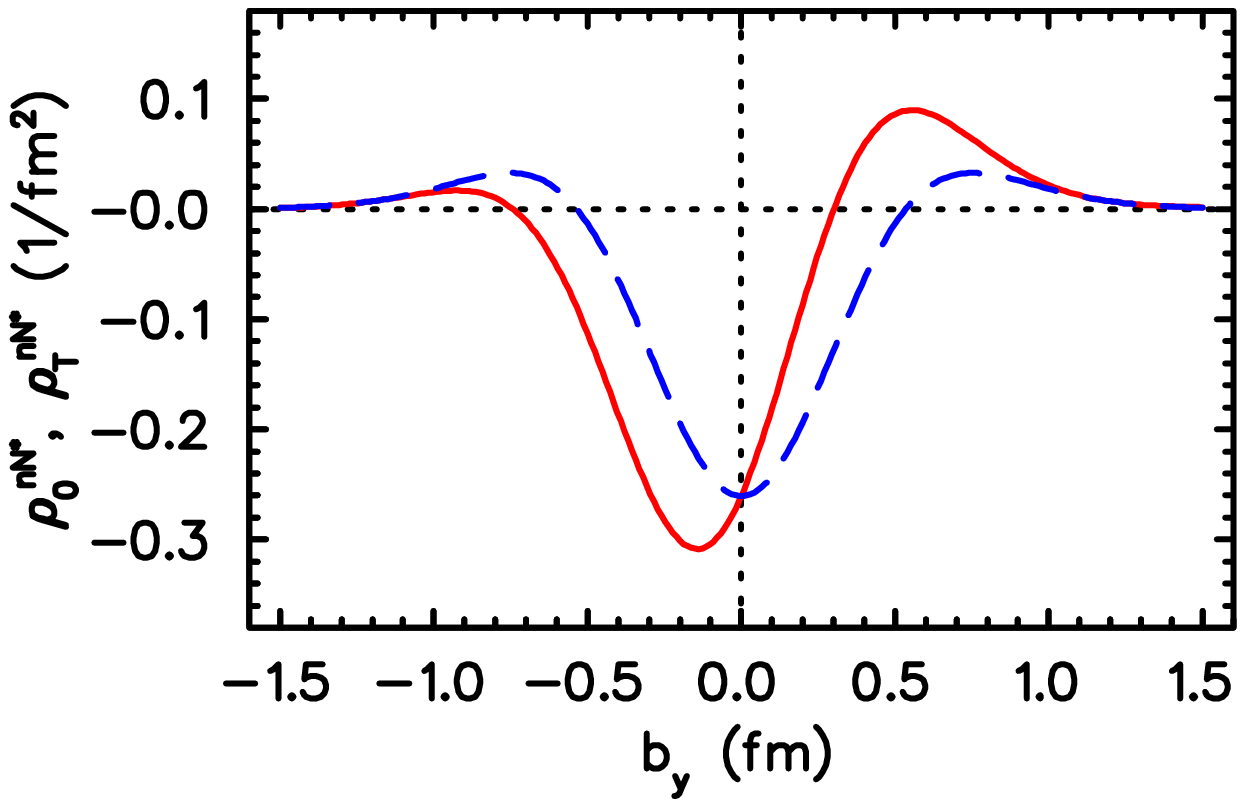}
\end{center}
\caption{Quark transverse charge density
corresponding to the $n \to P_{11}(1440)$ e.m. transition.
Upper left panel : when $n$ and $N^\ast$ are
unpolarized ($\rho_0^{n N^\ast}$).
Upper right panel : when $n$ and $N^\ast$ are polarized along the $x$-axis
($\rho_T^{n N^\ast}$). 
The light (dark) regions correspond with positive (negative) densities. 
Lower panel : densities
$\rho_T^{n N^\ast}$ (solid curves) and
$\rho_0^{n N^\ast}$ (dashed curves) along the $y$-axis.
For the $n \to P_{11}(1440)$ e.m. transition FFs,
we use the MAID2007 fit~\cite{MAID2007}. }
\label{fig:nnstar5}
\end{figure}

We show the results for the $N \to P_{11}(1440)$ transition charge
densities both for the unpolarized case and for the case of
transverse polarization in Fig.~\ref{fig:nnstar4} for the proton and
in Fig~\ref{fig:nnstar5} for the neutron. We use the empirical
information on the $N \to N^\ast(1440)$ transition FFs as
parameterized in Table~(\ref{table:maidfitA}, \ref{table:maidfitS})
and shown in Fig.~\ref{fig:nnstar3}.

It is seen from Fig.~\ref{fig:nnstar4} that for the transition on a
proton, which is well constrained by data, there is an inner region
of positive quark charge concentrated within 0.5~fm, accompanied by
a relatively broad band of negative charge extending out to about
1~fm. When polarizing the baryon in the transverse plane, the large
value of the magnetic transition strength at the real photon point,
yields a sizeable shift of the charge distribution, inducing an
electric dipole moment. For the neutron, which is not very well
constrained by data, the MAID2007 analysis yields charge
distributions of opposite sign compared to the proton, with
active quarks spreading out over even larger spatial distances, see
Fig.~\ref{fig:nnstar5}.

It may be of interest to also extract these densities within baryon
structure models, and check that within such models the quarks
active in the e.m. transition from the nucleon to its first radial
excited state are spatially more spread out than e.g. is the case
for the e.m. $N \to \Delta$ transition.

In summary, we analyzed in the present work the e.m. $N \to P_{11}
(1440)$ transition based on recent data for the proton which extend
up to 5~GeV$^2$. We extracted both the helicity amplitudes as well
as the transition form factors. The latter were used to extract the
quark transverse charge densities inducing this transition. For the
proton, it was found that this transition from the nucleon to its
first radially excited state is dominated by up quarks in a central
region of around 0.5~fm and by down quarks in an outer band which
extends up to about 1~fm. We leave it as a topic for future work to
extend the present analysis to also map out the quark charge
densities for the $N \to S_{11}(1535)$ and $N \to D_{13}(1520)$
transitions, which have also been studied extensively in experiment.

\begin{acknowledgments}
The work of M.~V. is supported in part by DOE grant
DE-FG02-04ER41302.
\end{acknowledgments}


\begin{thebibliography}{99}

\bibitem{Pascalutsa:2006up}
  V.~Pascalutsa, M.~Vanderhaeghen, and S.~N.~Yang,
  Phys.\ Rept.\  {\bf 437}, 125 (2007).

\bibitem{Burkert:2004sk}
  V.~D.~Burkert and T.~S.~H.~Lee,
  Int.\ J.\ Mod.\ Phys.\  E {\bf 13}, 1035 (2004).

\bibitem{MAID2003}
  L.~Tiator, D.~Drechsel, S.~Kamalov, M.~M.~Giannini, E.~Santopinto and A.~Vassallo,
  Eur.\ Phys.\ J.\  A {\bf 19}, 55 (2004).

\bibitem{MAID2007}
  D.~Drechsel, S.~S.~Kamalov and L.~Tiator,
  Eur.\ Phys.\ J.\  A {\bf 34}, 69 (2007).

\bibitem{Aznauryan:2004jd}
  I.~G.~Aznauryan, V.~D.~Burkert, H.~Egiyan, K.~Joo, R.~Minehart and L.~C.~Smith,
  Phys.\ Rev.\  C {\bf 71}, 015201 (2005).

\bibitem{Aznauryan:2008pe}
  I.~G.~Aznauryan {\it et al.}  [CLAS Collaboration],
  Phys.\ Rev.\  C {\bf 78}, 045209 (2008).

\bibitem{Denizli:2007tq}
  H.~Denizli {\it et al.}  [CLAS Collaboration],
  Phys.\ Rev.\  C {\bf 76}, 015204 (2007).

\bibitem{Joo:2005gs}
  K.~Joo {\it et al.}  [CLAS Collaboration],
  Phys.\ Rev.\  C {\bf 72}, 058202 (2005).

\bibitem{Biselli:2008zz}
  A.~S.~Biselli {\it et al.},
  Phys.\ Rev.\  C {\bf 78}, 045204 (2008).

\bibitem{Joo:2004mi}
  K.~Joo {\it et al.}  [CLAS Collaboration],
  Phys.\ Rev.\  C {\bf 70}, 042201 (2004).

\bibitem{Egiyan:2006ks}
  H.~Egiyan {\it et al.}  [CLAS Collaboration],
  Phys.\ Rev.\  C {\bf 73}, 025204 (2006).

\bibitem{Park:2007tn}
  K.~Park {\it et al.}  [CLAS Collaboration],
  Phys.\ Rev.\  C {\bf 77}, 015208 (2008).

\bibitem{Armstrong:1998wg}
  C.~S.~Armstrong {\it et al.}  [Jefferson Lab E94014 Collaboration],
  Phys.\ Rev.\  D {\bf 60}, 052004 (1999).

\bibitem{Thompson:2000by}
  R.~Thompson {\it et al.}  [CLAS Collaboration],
  Phys.\ Rev.\ Lett.\  {\bf 86}, 1702 (2001).

\bibitem{Merkel:2007ig}
  H.~Merkel {\it et al.}  [A1 Collaboration],
  Phys.\ Rev.\ Lett.\  {\bf 99}, 132301 (2007).

\bibitem{Alexandrou:2007dt}
  C.~Alexandrou, G.~Koutsou, H.~Neff, J.~W.~Negele, W.~Schroers and A.~Tsapalis,
  Phys.\ Rev.\  D {\bf 77}, 085012 (2008).

\bibitem{Lin:2008qv}
  H.~W.~Lin, S.~D.~Cohen, R.~G.~Edwards and D.~G.~Richards,
  arXiv:0803.3020 [hep-lat].

\bibitem{Miller:2007uy}
  G.~A.~Miller,
  Phys.\ Rev.\ Lett.\  {\bf 99}, 112001 (2007).

\bibitem{Carlson:2007xd}
  C.~E.~Carlson and M.~Vanderhaeghen,
  Phys.\ Rev.\ Lett.\  {\bf 100}, 032004 (2008).

\bibitem{Carlson:2008zc}
  C.~E.~Carlson and M.~Vanderhaeghen,
  arXiv:0807.4537 [hep-ph].

\bibitem{Alexandrou:2008bn}
  C.~Alexandrou {\it et al.},
  arXiv:0810.3976 [hep-lat].

\bibitem{Frolov:1998pw}
  V.~V.~Frolov {\it et al.},
  Phys.\ Rev.\ Lett.\  {\bf 82}, 45 (1999).

\bibitem{Ungaro:2006df}
  M.~Ungaro {\it et al.}  [CLAS Collaboration],
  Phys.\ Rev.\ Lett.\  {\bf 97}, 112003 (2006).

\bibitem{SAID}
  R.A.~Arndt, W.J.~Briscoe, I.I.~Strakovsky, R.L.~Workman,
  http://gwdac.phys.gwu.edu/.

\bibitem{Amsler:2008zz}
  C.~Amsler {\it et al.}  [Particle Data Group],
  Phys.\ Lett.\  B {\bf 667}, 1 (2008).

\end{thebibliography}
\end{document}